# Investigation of Cosmic-Ray Sources with Gamma-Ray Initiated Showers

**A.V. Uryson[1]**
*Lebedev Physical Institute, RAS*
*Leninsky pr. 53, Moscow 119099, Russia*
*E-mail:* `uryson@gmail.com`

A new method of investigating ultra-high energy cosmic ray sources is suggested. The method is based on analysis of gamma-ray emission that is generated in extragalactic space when ultra-high energy cosmic particles interact with cosmic background. We have found that intensity of the gamma-ray emission depends on characteristics of cosmic ray sources, specifically on their remoteness and initial particle energy spectra. In the Earth atmosphere cosmic rays initiate air showers, therefore selecting quanta-initiated showers (and excluding those from the galactic plane, gamma-ray sources, etc.) we can obtain above mentioned source characteristics. We derive that the number of quanta-initiated showers is 0 or ~1000 depending on source parameters, typical statistics of showers registered at $10^{14}$ eV being of ~$10^8$. The difference is large enough to use this method for studying ultra-high energy cosmic ray sources.

---

[1]Speaker

**1.    Introduction**

The main problem of ultra-high energy (UHE) cosmic rays (CRs) is where they come from. Point UHECR sources seem to be appropriate but they are not discovered.

In principle it is possible to identify CR sources by arrival directions of registered particles. This approach is complicated by several factors. The first is angular resolution of arrays: being of $1^0$ at best, there are at least several extragalactic objects in the error-boxes around the line of arrival direction. Farer is the possible CR source more objects are in the error-box. To find the source among a number of various objects, model assumptions are required about particle acceleration in sources along with probabilistic analyzes. The second complicating factor is insufficient information about extragalactic magnetic fields. Charged particles are deviated by magnetic fields but the deflection is unknown. Magnetic fields are assumed to be weak enough providing CR rectilinear propagation in the space. Observational data and theoretical results [1, 2] indicate that this assumption is plausible in average. However some particles possibly are deflected at angles exceeding the angular resolution of array. In this case source identification using particle arrival directions is incorrect. We used previously this method and found that CRs were accelerated in jets of active galactic nuclei, assuming particles moved toward jet [3]. Treating this problem we reveal an additional factor that makes difficult source identification. Commonly jets are bended, therefore lines of CR arrival directions show toward probable sources not in every case.

Information about UHECR origin usually is obtained from particle energy spectrum. In space, particles lose energy in interaction with cosmic microwave background. This results in a lack of particles at $E>10^{20}$ eV at the Earth and in spectrum suppression (GZK-effect: [4, 5]) if UHECRs come from distances of more than ~100 Mpc. If particles arrive from nearer distances the UHECR spectrum has no GZK-suppression. Experimental data obtained with CR giant arrays - Pierre Auger Observatory and Telescope Array - indicate that UHECR energy spectrum is suppressed. However, there may be another reason for spectrum suppression besides GZK-effect: particles are accelerated to the maximum energy of ~$10^{21}$ eV due to conditions in possible UHECR sources [6, 7]. As sources are still unknown, the acceleration energy limit cannot be ruled out. So the reason for the spectrum suppression is not clear.

In addition to GZK-suppression there is another effect of UHECR interaction with cosmic microwave background. Particles initiate electromagnetic cascades in extragalactic space [8, 9]. We calculate intensity of cascade quanta at the energy $E \approx 10^{14}$ eV and find that it depends on source characteristics: remoteness and initial CR energy spectra. We chose this energy because the Universe is almost opaque for such quanta [10], so their intensity is maximal.

Gamma-ray emission at $E \approx 10^{14}$ eV is generated in other processes as well. First, it is produced in the Galaxy in CR interaction with interstellar gas, CR energy being $E<10^{15}$ eV. Second, pulsars emit gamma-quanta at $10^{14}$ eV. Besides, active galactic nuclei radiate quanta with the energy $10^{14}$ eV. Finally, many gamma-ray sources seems to be unresolved. These sources contribute to gamma-ray background. How can we select quanta generated in cascades from quanta of other origin? For the first process, the intensity of gamma-ray emission reaches its maximum in the galactic plane. Discrete sources being gamma-ray emitters, there are peaks in the intensity toward emitters. As for contribution of unresolved sources it is estimated theoretically. Using these data it is possible to select cascade gamma-quanta from those of other origin.

Penetrating in the Earth atmosphere quanta initiate external air showers that are registered with ground arrays. Selecting these showers from all showers registered (and excluding quanta-initiated showers arrived from the galactic plane, gamma-ray sources, etc.) it is possible to study showers initiated by cascade quanta.

The total statistics of showers detected with a ground array typically is $\sim 10^8$ events. We show that this statistics is enough to determine parameters of CR sources analyzing cascade emission.

## 2. Extragalactic cascades

In extragalactic space UHECRs interact with microwave background: $p+\gamma_{rel} \to p+\pi^0$, $p+\gamma_{rel} \to n+\pi^+$. Cross section of reactions depends on energies of interacting particles. It has a peak of approximately 0.44 mb at photon energy $\varepsilon^* \approx 340$ GeV in the proton rest system [8]. Thus CR free path depends on particle energy. CRs being at energies of about $4 \cdot 10^{19}$ eV their paths are ~100 Mpc, whereas particles at energies approximately $10^{20}$ eV have paths of only ~10 Mpc [9]. The energy transfer to pions is determined by the inelasticity of the interaction that is given also in [12].

Particles cascading is described in [11, 12]. Generated pions decay via reactions $\pi^0 \to 2\gamma$, $\pi^+ \to \mu^+ + \nu$, $\mu^+ \to e^+ + \nu + \nu$, producing electrons and quanta which give rise to cascades. Then cascade develops through electron and quanta reactions with background emission including relic and radio photons [10,13]. These reactions are pair production (PP) $\gamma + \gamma_b \to e^+ + e^-$ and inverse Compton scattering (ICS) $e + \gamma_b \to e' + \gamma'$ (here index $b$ denotes background photons, index ′ denotes scattered particles). PP takes place provided that the photon energy is above the threshold energy

$$E_t = (mc^2)^2 / \varepsilon_b, \qquad (1)$$

where $mc^2$ is the electron mass, $mc^2 = 0.511$ MeV, and $\varepsilon_b$ is the energy of the background photon.

Cross sections of the PP and ICS are known [14]. For PP it is

$$\sigma_{\gamma\gamma} \approx 3/8 \sigma_T a^2 / \{[2+2a^2-a^4] \ln(a^{-1}+(a^{-2}-1)^{1/2}) - (1-a^2)^{1/2}(1+a^2)\}, \qquad (2)$$

where $a = mc^2/E_{ec}$, $E_{ec}$ is the photon energy in the centre-of-mass system,

$$E_{ec} = [E_e \varepsilon_b (1-\cos\psi)]^{1/2}, \qquad (3)$$

and $\psi$ is the angle between photon impulses in the laboratory system. The cross section of the ICS equals at $E_e > E_t$

$$\sigma_{IC} \approx 3/8 \sigma_T (mc^2)^2 / (E_e \varepsilon_b) \{\ln[2 E_e \varepsilon_b / (mc^2)^2] + 0.5\} \qquad (4)$$

and equals Thomson cross section, $\sigma_T \approx 6.65 \cdot 10^{-25}$ cm$^2$ at $E_e < E_t$.

Electrons or quanta being at lower energies, $E_e < E_t$, there is no PP, and comparatively soft quanta are produced in ICS, at mean energies

$$E_\gamma = 4/3 \varepsilon_b (E_e/mc^2)^2. \qquad (5)$$

In this case electron loses energy fairly slowly.

There are two other interactions of quanta and electron with background photons: double PP $\gamma + \gamma_b \to e^+ + e^- + e^+ + e^-$ and $e + \gamma_b \to e' + e^+ + e^-$, but their contributions to cascade are negligible [15].

Electrons produce photons also via synchrotron radiation in magnetic fields where electrons lose energy emitting fairly soft photons. Depending on field value synchrotron radiation can be intensive enough to stop cascade developing. Luckily, extragalactic magnetic fields are weak to violate cascade. We estimate the field value in [16] providing energy losses in synchrotron radiation are smaller than those in ICS. Here we use the following approach.

We assume that electrons scatter photons faster than they lose a half of energy in synchrotron

emission:

$$\Delta t_{IC} < T_s, \qquad (6)$$

where $\Delta t_{IC}$ is duration of IC scattering and $T_s$ is duration of synchrotron emission in which electron energy reduces in two times. The value of $T_s$ is given in [17]:

$$T_s = 5 \cdot 10^8 \, mc^2/(B_\perp^2 E_e), \qquad (7)$$

here $B_\perp$ in G is the field component normal to electron velocity, and the electron energy $E_e$ is in eV. The value $\Delta t_{IC}$ equals

$$\Delta t_{IC} = \lambda_{IC}/c, \qquad (8)$$

where $\lambda_{IC}$ is electron mean free path, $\lambda_{IC} = (\sigma_{IC} \cdot n_b)^{-1}$, $n_b$ is the background photon density, and $c$ is the speed of light.

Both approaches give that UHE electrons have minor synchrotron losses in the magnetic field $B \approx 10^{-11}$ G. For electrons at $E \approx 10^{14} - 10^{17}$ eV, losses are minor in the field $B < 2 \cdot 10^{-8}$ G. Finally, electrons at $E \approx 10^{14}$ eV have negligible synchrotron losses in the magnetic field $B < 5 \cdot 10^{-6}$ G. (The energy value of $10^{14}$ eV is the lowest energy which we assume.)

Study of extragalactic magnetic fields gives following results. Fields are inhomogeneous [1, 2]: there are regions with weak fields, $1 \cdot 10^{-17}$ G $< B < 3 \cdot 10^{-14}$ G, along with filaments where $B \approx 10^{-9} - 10^{-8}$ G in the extragalactic space. In both cases magnetic fields do not prevent cascade.

In the Galaxy, the magnetic field is $B \approx 3 \cdot 10^{-6}$ G. Here UHE electrons lose energy on synchrotron radiation faster than on cascade processes. So we assume no cascades in the Galaxy.

Cosmic particles move in the expanding Universe and adiabatically lose energy. Passing the distance from a point with red shift $z_i$ to a point with red shift $z_{i+1}$ the particle energy decrease $(\Delta E)_{ad}$ equals: $(\Delta E)_{ad} = E(z_{i+1} - z_1)/(1 + z_i)$.

## 3. The model

Model assumptions concern three points: CR sources, background emission in the extragalactic space, and the extragalactic magnetic field.

We assume that CR sources are point. There is no contradiction with results [18].

Next we assume that UHECR sources are active galactic nuclei: Blue Lacertae objects (Bl Lac's) that are powerful and remote at distances of hundreds Mpc (with red shifts up to ($z$=1.1), and Seyfert nuclei located in the radius of about 50 Mpc (with red shifts $0 \leq z \leq 0.0092$) which are comparatively low-power. These possible UHECR sources were analyzed in [19-21].

We obtained spatial distributions of sources using data on object red shifts from the catalogue [22]. The distributions are presented in [23].

We suppose that at $z > 0.1$ distance $r$ is related to red shift z as

$$r = 2/3 c H^{-1}(1 - (1+z)^{-3/2}) \text{ (Mpc)} \qquad (9)$$

(in accordance with Einstein-de Citter model with the parameter $\Omega = 1$), and as

$$r = cz/H \text{ (Mpc)} \qquad (10)$$

at lower $z$.

We suppose a simple case, when in sources CRs are accelerated either by electrical fields, or by shock waves. In the first case the initial CR spectrum is monoenergetic, in the second one it is exponential [23]. To describe UHECR energy spectra measured at the Earth the particle spectrum in Bl Lacs is either monoenergetic with the energy $E = 10^{21}$ eV or exponential with the index in the range 2 – 3; in Seyfert nuclei CR spectrum is exponential with the index in the range 2 – 3. Particles in sources are accelerated to the energy not higher than $10^{21}$ eV [6, 7].

Herein after Bl Lacs are called "remote" sources, Seyferts are called "near" ones.

Thus model assumptions on UHECR sources are distances to them and initial CR spectra. In total we consider three types of UHECR sources: "remote" ones with the particle monoenergy spectrum, "remote" sources with the CR exponential spectrum, and "near" ones with the particle exponential spectrum.

We assume also that CRs consist of protons.

In the model, background emission is treated in the following way. Microwave background emission has Planck distribution in energy with the mean value $\varepsilon_r = 6.7 \cdot 10^{-4}$ eV, the mean photon density is $n_r = 400$ cm$^{-3}$. High-energy "tail" consists of photons at the energy $\varepsilon_t = 1 \cdot 10^{-3}$ eV, their mean density is $n_t = 42$ cm$^{-3}$. The threshold energy (1) of PP and ICS equals $E_t \approx 10^{14}$ eV. Microwave background anisotropy has no influence on cascade and it is not considered.

Spectrum of the extragalactic background radiation was obtained experimentally in [24] and later theoretically in [25] allowing for evolution of radio sources. We use results [25]: radio background exists at the energy $\varepsilon_{rad} \approx 4 \cdot 10^{-10}$ eV, so the threshold energy (1) equals $E_t \approx 6 \cdot 10^{20}$ eV.

We neglect electron losses due to synchrotron radiation in extragalactic magnetic fields.

## 4. Calculations

We determine the number of gamma-quanta at the energy $E_\gamma \approx 10^{14}$ eV produced by a CR proton propagating from the source to the Galaxy by the following scheme.

Distances from UHECR sources are obtained in accordance with their spatial distribution: a source red shift $z_s$ is generated by Monte-Carlo technique, and the distance to the source is calculated. Proton mean free path in interaction with background emission being $<L> = 1/(n_{rel}\sigma_{p\gamma})$, the proton free path $L$ is generated by Monte-Carlo method. In the point of interaction proton energy losses $(\Delta E)_{ad}$ is calculated due to the expansion of the Universe. Then the energy of CR proton after interaction is obtained. Both the cross-section of interaction and energy transfer are calculated using their dependences on $\varepsilon^*$ [12]. The procedure is repeated until proton reaches the Galaxy either its energy decreases to $E < 4 \cdot 10^{19}$ eV (because protons at this energy have paths of several hundred Mpc).

In $\pi^0$-decays quanta are produced at the energy $E_\gamma = 0.5 E_{\pi 0}$. In PP according to (1-3), the angle $\psi$ included between impulses of a cascade photon and a background one is generated in the observer's frame of reference, then the photon energy $E_{ec}$ in the centre of mass system is calculated. Next we obtain values of the cross section $\sigma_{\gamma\gamma}$ and the free path of a cascade quanta $\lambda_{\gamma\gamma} = 1/(n_b\sigma_{\gamma\gamma})$. We chose between radio photon and relic one that photon which path is shorter. Then the threshold energy $E_t$ is derived according to (1). We suppose that one of the particles in the electron-positron pair has the energy $E_t$ and the other one is at the energy $(E_\gamma - E_t)$. The ICS is considered in the same way using (4, 5). Electron is treated until its energy decreases to $10^{14}$ eV (as at lower energies it loses energy fairly slowly scattering soft quanta).

Using this scheme we obtain the total number of gamma-quanta $N_\gamma$ at the energy $E_\gamma \approx 10^{14}$ eV produced by one proton passing from a source to the Galaxy.

The procedure is repeated for 1000 sources emitting one proton. Then the cascade-mean number of gamma-quanta at the energy $E_\gamma \approx 10^{14}$ eV is derived.

## 5. Results

In the model of "remote" sources with the CR monoenergetic initial spectrum, $E = 10^{21}$ eV, protons on their way to the Galaxy give rise to $N_\gamma \approx 10^6$ gamma-quanta at the energy $E_\gamma \approx 10^{14}$ eV.

Sources with CR exponential initial energy spectra result in smaller values of $N_\gamma$. Protons from "remote" sources initiate $N_\gamma \approx 1.5 \cdot 10^3$ quanta at $E_\gamma \approx 10^{14}$ eV along their way to the Galaxy. "Near" sources (which are Seyfert nuclei) result in $N_\gamma \approx 0$ quanta reaching the Galaxy at $E_\gamma \approx 10^{14}$. These values of $N_\gamma$ do not depend on the exponential index of the initial energy spectrum.

## 6. Discussion

Variety in the number of cascade gamma-quanta $N_\gamma$ arises from two reasons: resonance behavior of the cross section of p-$\gamma_{rel}$-interaction and initial CR spectrum. The emitted proton interacts with microwave background radiation until its energy decreases to $4 \cdot 10^{19}$ eV. Then its free path grows to several hundred Mpc and the proton has a poor chance of interacting with background emission along the rest of the way to the Earth. As a result a proton from sources with the monoenergetic energy spectrum, the initial energy being of $10^{21}$ eV, interacts with cosmic microwave background about 10 times, initiating approximately 10 electromagnetic cascades. In sources with CR initial exponential energy spectra the majority of protons are emitted at the energy about $5 \cdot 10^{19}$ eV regardless of the exponential index of the spectrum. Having long free paths protons interact with background emission 1-2 times only, thus giving rise to 1-2 cascades. In addition, distances from "near" sources are too short for cascade developing. The proton mean path as a function of the proton energy is shown in Fig. 1.

The quantum mean path in PP as a function of quantum energy is shown in Fig.2, and electron mean path in ICS versus electron energy is shown in Fig.3. Paths do not exceed 10 Mpc for all quantum/electron energies. Quantum/electron propagation has no influence on the variety in the number $N_\gamma$ in models considered.

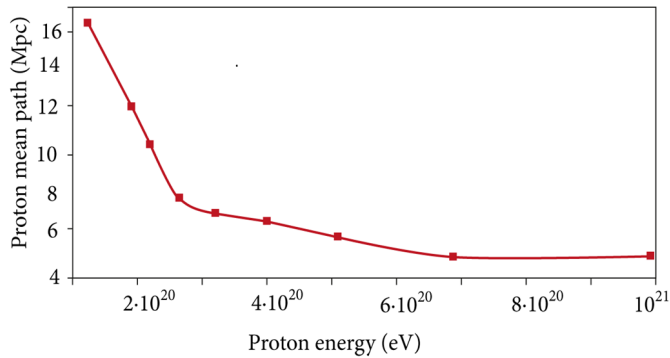

Fig.1. Proton mean path vs. proton energy in extragalactic space

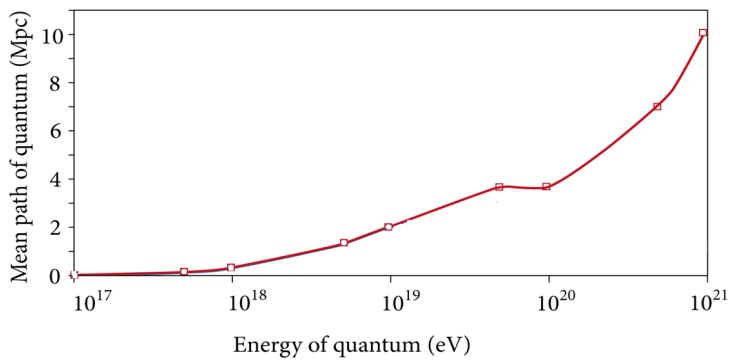

Fig.2. Quantum mean path vs. its energy in PP in extragalactic space

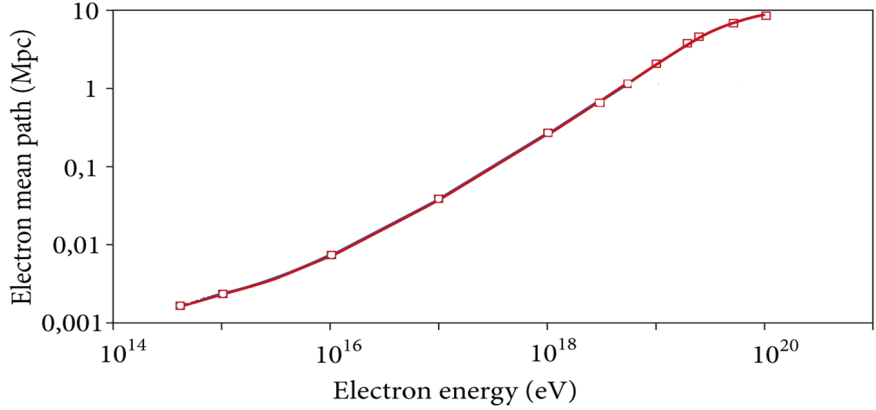

Fig.3. Electron mean path vs. its energy in ICS in extragalactic space

The fraction of cascade quanta-initiated showers at $E_\gamma \approx 10^{14}$ eV relatively to the total number of showers at this energy equals

$$f = I_\gamma(10^{14}\text{ eV})/I_{CR}(10^{14}\text{ eV}) = N_\gamma I_{CR}(E,\text{ eV})/I_{CR}(10^{14}\text{ eV}), \quad (11)$$

where $E$ is the energy of cosmic proton that initiated cascades, $I_{CR}$ is the CR intensity.

We determine CR intensity $I_{CR}$ at energy $E$ and at $10^{14}$ eV supposing $I_{CR}(E) = I_{CR}(>E)$. It is valid because CR spectra are steep. Approximations of CR integral intensity are [26]:

$$I_{CR}(>E) = 1 \cdot (E\text{ (GeV)})^{-1.7} \text{ particle}/(\text{cm}^2\cdot\text{c}\cdot\text{cr}) \quad (12)$$

at $10\text{ GeV} < E < 3\cdot 10^6$ GeV;

$$I_{CR}(>E) = 3\cdot 10^{-10} (E\text{ (GeV)}/10^6)^{-2.1} \text{ particle}/(\text{cm}^2\cdot\text{c}\cdot\text{cr}) \quad (13)$$

at $E > 3\cdot 10^6$ GeV.

In our models the fraction of cascade quanta-initiated showers is the following. "Remote" and powerful sources with the initial monoenergetic energy spectra, $E = 10^{21}$ eV:

$$f \approx 3\cdot 10^{-5}\ (N_\gamma \approx 10^6);$$

"remote" sources with the initial exponential energy spectra:

$$f \approx 1.3\cdot 10^{-13}\ (N_\gamma \approx 1.5\cdot 10^3);$$

"near" sources with the initial exponential energy spectra:

cascades have no time to develop, and $f = 0$ ($N_\gamma \approx 0$).

Typically the statistics of showers at $E \approx 10^{14}$ eV registered with a ground array is $N_{shower} \sim 10^8$. The number of cascade-quanta-initiated showers $N_{gamma}$ among them equals: $N_{gamma} \approx 3\cdot 10^3$ if UHECR sources are "remote" and powerful, $N_{gamma} \approx 0$ if energy spectra in sources are exponential regardless of distances to sources and of the index of the spectrum.

## 7. Conclusion

We analyzed three types of possible UHECR sources: "remote" and powerful sources located at hundreds Mpc with red shifts up to $z = 1.1$, having the initial monoenergetic spectrum, $E = 10^{21}$ eV; "remote" sources but with the initial exponential energy spectra, and finally "near" sources located in the radius ~50 Mpc, with red shifts $0 \leq z \leq 0.0092$, also having exponential spectra. Source parameters – remoteness and initial energy spectrum - were chosen following [23]. It appears that at the Earth the fraction of cascade quanta-initiated showers at $10^{14}$ eV strongly depends on source parameters.

Typically the statistics of showers at $E \approx 10^{14}$ eV registered with a ground array is $N_{shower} \sim 10^8$. Then the number of cascade-quanta-initiated showers is $N_{gamma} \approx 3\cdot 10^3$ if UHECR sources are "remote" and powerful. $N_{gamma} \approx 0$ if sources have initial exponential energy spectra regardless of distances to sources and of the index of the spectrum. In reality there can be a mixture of powerful "remote" sources with other sources. Then $N_{gamma}$ is less than $3 \cdot 10^3$.

Selection of cascade quanta-initiated showers at $E \approx 10^{14}$ eV can be performed using data of GAMMA, KASCADE, and other arrays.